\documentclass[12pt]{article}
\usepackage{epsfig,amsfonts,amssymb}
\usepackage{hyperref}
\pdfoutput=1

\usepackage{cite}
\usepackage{cite}

\usepackage{comment}

\def\ZZZ{{\hbox{ Z\kern-1.6mm Z}}}
\def\RRR{{\hbox{ R\kern-2.4mm R}}}
\def\CCC{{\hbox{ C\kern-2.0mm C}}}
\def\zzz{{\hbox{z\kern-1mm z}}}

\def\ZZZ{\mathbb{Z}}
\def\RRR{\mathbb{R}}

\newcommand{\qeq}{{\hbox{=\kern-2.3mm ? \kern.5mm }}}
\renewcommand{\qeq}{=}

\newcommand{\eps}{\epsilon}

\newcommand{\GG}{{\cal G}}

\newcommand{\EE}{{\cal E}}

\newcommand{\NN}{{\cal N}}

\newcommand{\be}{\begin{equation}}
\newcommand{\ee}{\end{equation}}
\newcommand{\ben}{\begin{eqnarray}\displaystyle}
\newcommand{\een}{\end{eqnarray}}

\newcommand{\refb}[1]{(\ref{#1})}
\newcommand{\p}{\partial}
\newcommand{\sectiono}[1]{\section{#1}\setcounter{equation}{0}}

\def\one{{\hbox{ 1\kern-.8mm l}}}
\def\zero{{\hbox{ 0\kern-1.5mm 0}}}

\newcommand{\bea}[1]{\begin{eqnarray}\label{#1} }
\newcommand{\eea}{\end{eqnarray}}

\newcommand{\eqref}{\refb}

\usepackage{bm}
\usepackage[table]{xcolor}

\begin{document}

\baselineskip 24pt

\begin{center}

%{\Large \bf How to Study Isolated Exotic Branes}
{\Large \bf Extended Supergravity Needs String Scale Cut-off}

\end{center}

\vskip .6cm
\medskip

\vspace*{4.0ex}

\baselineskip=18pt

\centerline{\large \rm Ashoke Sen}

\vspace*{4.0ex}

\centerline{\large \it International Centre for Theoretical Sciences - TIFR 
}
\centerline{\large \it  Bengaluru - 560089, India}

%\centerline{\large \it ~$^c$Homi Bhabha National Institute}
%\centerline{\large \it Training School Complex, Anushakti Nagar,
%    Mumbai 400085, India}

\vspace*{1.0ex}
\centerline{\small E-mail:  ashoke.sen@icts.res.in}

\vspace*{5.0ex}

\centerline{\bf Abstract} \bigskip

Many string compactifications down to four non-compact space-time 
dimensions with $\mathcal N=8$,  $\mathcal N=6$ and $\mathcal N=4$ 
supersymmetry
have BPS black holes carrying pure D-brane charges and preserving four supersymmetries.
The string coupling does not flow in these backgrounds and can be set to any arbitrary value. Therefore
the supersymmetric index of these black holes must be independent of the string coupling. On the other
hand, explicit
computation of one loop correction to the index from gravitational path integral
is sensitive to the choice of ultraviolet cut-off. We show that
if the cut-off scale is chosen to be the string scale in accordance with the rules of string theory, then the
dependence of the index on the string coupling disappears in accordance with the expectation from
supersymmetry. 
Similar results are obtained for type II string theories compactified on Calabi-Yau manifolds with 
zero Euler number. For non-zero Euler number we  encounter a puzzle that we discuss but do not 
fully resolve.

\vfill \eject

\tableofcontents

\section{Introduction} \label{s1}

Gravitational theories with extended supersymmetry in four space-time dimensions have BPS states
carrying U(1) gauge charges. Those that preserve only four supersymmetries are particularly important since
gravitational backreaction turns them into charged black holes when the 
gravitational coupling is sufficiently strong. 
For these black holes we can define an appropriate index, 
which is protected against the change in the coupling constant $g_s$
as a consequence of supersymmetry. On
the other hand, we can compute this index using gravitational path integral and test if it is really
protected against the change in $g_s$. While at tree level supergravity this result follows
from the so called attractor mechanism\cite{9508072,0506177}, this is not obvious at the loop level.  In fact at the
loop level the result is sensitive to the ultraviolet cut-off of the theory. The goal of this paper will be
to show that if the ultraviolet cut-off is chosen to be the string scale then the one loop correction to the
index indeed becomes independent of the string coupling in $\mathcal N=8$, $\NN=6$ and $\mathcal N=4$ 
supersymmetric theories
and a class of $\mathcal N=2$ supersymmetric theories, obtained from type II string theory compactification
on Calabi-Yau manifolds with vanishing Euler number. 
On the other hand, for different choice of cut-off, the index in the $\NN=8$, $\NN=6$ and $\NN=2$ supersymmetric
theories become $g_s$ dependent, contradicting the prediction of 
supersymmetry.
Hence for these theories, maintaining supersymmetry
at the loop level seems to necessitate the use of string theory as the ultraviolet completion of the theory.

This result may seem surprising since {\it a priori} supergravity does not know anything about the
string scale. The way the information about the string scale enters our analysis is through the charge
quantization rules. If
the various gauge fields are normalized so that they couple to charges whose quanta are of order unity,
independent of the asymptotic values of the moduli fields,
then the bosonic part of the classical action $S$ scales as $\sigma^2$ if we shift the dilaton by
$-\ln\sigma$ and scale the Ramond Ramond sector fields by $\sigma$, leaving fixed the
string frame metric and other NSNS sector fields. This scaling law implicitly
introduces the dilaton and the string scale (up to a constant multiplicative factor).
This will be discussed in more detail in section \ref{s4}.

We now describe the set-up we shall be studying. We consider four kinds of theories:
\begin{enumerate}
\item Type IIA and IIB string theories with asymptotic geometry $T^6\times R^{3,1}$. These theories
have $\mathcal N=8$ supersymmetry.
\item Type IIA string theory with asymptotic geometry
$T^4\times T^2\times R^{3,1}$, orbifolded by a $\ZZZ_2$ transformation that acts as
a translation along $T^2$ and a
T-duality along all the circles
of $T^4$.
This theory has $\NN=6$ supersymmetry in $D=4$\cite{9508064}.
\item Type IIA and IIB string theories  with asymptotic geometry $K3\times T^2\times R^{3,1}$, 
or some of its orbifolds where the orbifold group acts as translation along a circle of $T^2$ and
a transformation on $K3$ that preserves all the supersymmetries of the original theory.
These theories have $\mathcal N=4$ supersymmetry and are the type II duals of the 
CHL models\cite{9505054}.
\item Type IIA and IIB string theories  with asymptotic geometry $CY_3\times R^{3,1}$, where
$CY_3$ is a Calabi-Yau threefold. 
\end{enumerate}
In all of these theories we consider black holes that carry only RR charges. Hence they are 
associated with  D-branes
wrapped on various compact cycles. The string coupling $g_s$ does not flow in the background of these
black holes and can be set to any constant value. In particular, by taking it to be small we can ensure
that perturbation theory is trustable. We can then compute the one loop contribution to the index and
check if it is independent of $g_s$.

The index that we compute is the helicity supertrace\cite{9611205,9708062}
\be \label{ebndef}
B_n(q,\beta) = {(-1)^{n/2}\over n!}\, Tr_{q,\vec k=0}[(-1)^F e^{-\beta H} (2h)^n]\, ,
\ee
where $Tr_{q,\vec k=0}$ denotes that the trace is taken over all states at rest, 
carrying fixed electric and magnetic U(1) charges collectively denoted as
$q$, and $h$ is the third component of the angular momentum in the rest frame. 
This counts, with appropriate sign, the number of BPS supermultiplets that break at
most $2n$ supersymmetries. Therefore to count states that preserve four supersymmetries,
we choose $2n=28$ for $\mathcal N=8$ supersymmetric theories, 
$2n=20$ for $\mathcal N=6$ supersymmetric theories, 
$2n=12$ for $\mathcal N=4$ supersymmetric theories
and $2n=4$ for $\mathcal N=2$ supersymmetric theories. If $e^{S_{BPS}}$ denotes this number, then
$B_{n}$ is related to $S_{BPS}$ via the relation
\be
B_{n}(q,\beta) = e^{S_{BPS}(q) - \beta \, M_{BPS}(q)}\, ,
\ee
where $M_{BPS}(q)$ is the mass of the BPS state, giving the eigenvalue of $H$ at $\vec k=0$.
Therefore we can compute $S_{BPS}$ using the relation
\be
S_{BPS} = \ln B_{n}(q,\beta) + \beta \, M_{BPS}(q)\, .
\ee

$B_{n}$ can be computed in terms of gravitational path 
integral\cite{Gibbons:1976ue} with appropriate 
twisted boundary condition at infinity that takes into account the insertion of $(-1)^F$ into the
trace in \refb{ebndef} 
\cite{1810.11442,2107.09062,2310.07763,2411.08260,2411.12413,2501.17909,2510.23699,2511.18771} 
and an explicit insertion of $(2h)^n$ 
to account for the $(2h)^n$ factor
in the trace\cite{2306.07322,2308.00038,2605.30417}. 
This path integral does not directly compute $B_n$, instead it includes trace
over all the momentum states and also over states carrying different  electric charges  weighted by
appropriate chemical potential. Nevertheless we can extract $B_n$ and $S_{BPS}$ from this
computation via appropriate inverse Laplace transform\cite{2308.00038,2605.30417}.

During our analysis we shall make heavy use of the results of \cite{2605.30417} on the results
of gravitational path integral evaluation of $S_{BPS}$. Ref.~\cite{2605.30417} analyzed $S_{BPS}$
for states that carry only NSNS electric and magnetic charges while here we are analyzing
states that carry only RR charges. Hence the set up is different. Nevertheless the results of 
\cite{2605.30417} are general enough to be adapted to the analysis of the current paper. For
this reason we devote section \ref{s2} reviewing the main results of \cite{2605.30417}, but write it in a
language that can be readily used in the analysis of the current paper.

In section \ref{s3} we use these results to compute the one loop correction to $S_{BPS}$ for the
states of interest. The result that we find is the following. For $\mathcal N=8$, $\NN=6$ 
and $\mathcal N=4$ supersymmetric
theories, the $g_s$ dependence of various contributions to $S_{BPS}$ cancel and we are left
with a $g_s$ independent $S_{BPS}$ as predicted by supersymmetry. However, for type IIA/IIB theory
compactified on a Calabi-Yau threefold, we get, 
\be
e^{S_{BPS}} \propto g_s^{\pm \chi/12}\, ,
\ee
where the $+/-$ sign holds for type IIB/IIA theories. This is in contradiction to the expected $g_s$
independence of $S_{BPS}$ that follows from supersymmetry. In section \ref{s5} we suggest
possible reasons for this but do not arrive at a definite 
conclusion.\footnote{As we shall discuss in section \ref{s5}, for $\NN=2$ theories with
$\chi\ne 0$, there may be other
reasons for the index to acquire dependence on the string coupling, or the path integral to
receive additional contributions, but these reasons are absent
in the $\NN=8$, $\NN=6$ and $\NN=4$ theories.}

In section \ref{s4} we explore what happens if we choose the cut-off scale to be some other scale.
In particular we consider the case where the cut-off length scale is $g_s^\gamma$ times the string scale,
-- {\it e.g.} $\gamma=1$ would correspond to choosing the Planck scale as the cut-off scale.
We find that while for $\NN=4$ theories the result for $S_{BPS}$ remains independent of $\gamma$,
for $\NN=8$, $\NN=6$ and $\NN=2$ theories the results for $S_{BPS}$ depends on $\gamma$ and for any
$\gamma\ne 0$, $S_{BPS}$ acquires additional $g_s$ dependence contradicting the expectations
from supersymmetry. This suggests that maintaining extended supersymmetry in the full quantum
theory requires using the string scale as the ultraviolet cut-off scale.

We shall end the introduction by giving a general formula from which we can derive the results of
\cite{2605.30417} as well as all the results of this paper. Suppose the ultraviolet length cut-off is given by $g_s^\gamma$
times the string length scale. Then the one loop logarithmic correction to the logarithm of the index is given by (see \refb{etotal}, \refb{ealphans} and \refb{egammadep})
\be
c_1 \, \ln\, \lambda_S + c_2 \, \ln\, g_s + c_3 \, \gamma\, \ln\,  g_s \, ,
\ee
where 
\be  c_1 = \cases{-8\quad \hbox{for $\NN$=8 theory} \, ,\cr
 -4 \quad \hbox{for $\NN$=6 theory} \, ,\cr
 0  \quad \hbox{for $\NN$=4 theories} \, , \cr
{1\over 6} \, (23 - n_V + n_H)
 \quad \hbox{for $\NN$=2 theories}
}
\ee
\be 
c_2 = n_{NS} -4, \qquad c_3=-c_1\, ,
\ee
wheer
$ n_{NS}$ denotes the number of U(1) gauge fields arising from the NSNS sector,
and, in the $\NN=2$ supersymmetric theory,
$n_V$ is the number of vector multiplets
and 
$n_H$ is the number of hypermultiplets. From the perspective of supergravity, the
NSNS and RR sector gauge fields are distinguished by whether or not the gauge field
kinetic term is multiplied by a factor of $e^{-2\phi}$ where $\phi$ is the dilaton field.

\sectiono{Collection of useful results} \label{s2}

We shall begin by collecting some useful results, all of which can be found in \cite{2605.30417}. All
length and mass scales in this discussion will be specified in string units unless mentioned otherwise.

We consider black holes preserving four supersymmetries in four dimensional $\mathcal N=8$, 6, 
4 and 2
supersymmetric theories. 
Once we have chosen a duality frame to describe the theory, we can divide the total set of
charges $q$ into two sets: the electric charges denoted by $Q$ and the magnetic charges denoted by
$P$. For the RR sector gauge fields 
this division is somewhat arbitrary, and the final result will not depend on
this choice. We adjust the asymptotic values of the scalars so that
all the scalars remain (more or less) constant
in the black hole geometry and curvature and other field strengths in the black hole background
are controlled by one overall length scale, which we call $\lambda_S$ when measured 
in the string scale.\footnote{In the analysis of \cite{2605.30417}, $\lambda_S$ was set by the
overall scale $\lambda_P$ of the NSNS sector magnetic charges and the string coupling $g_s$ was set by the
square root of the ratio of the NSNS sector magnetic and electric charges. The black holes
we shall consider will be different and hence these properties will no longer be valid. Nevertheless
we can use the results of \cite{2605.30417} written in terms of $\lambda_S$ and $g_s$, with the
understanding the $\lambda_S$ is the size of the black hole measured in the string scale and $g_s$
is the near horizon value of the string coupling.}
We also denote by $S_{BPS}$ the logarithm of the appropriate helicity
trace index that
counts the number of black hole microstates with sign and by $M_{BPS}$ the mass of the 
supersymmetric black hole of interest.
Let $Z_n$ be the partition function computed 
using gravitational path
integral with the insertion of $(2h)^n$ where $h$ is the Noether charge measuring the third
component of the angular momentum and $2n$ is the number of supersymmetries broken by the
black hole. The boundary conditions on the fields are chosen so that the 
inverse temperature $\beta$ given by the period of the Euclidean time circle is of order
$\lambda_S$ in string units, the chemical potential $\omega$
dual to the third component of the angular momentum has the value $2\pi i/\beta$ and the 
chemical potentials dual to various electric charges are such that the rotating black hole saddle that
gives the dominant contribution to the path integral carries the desired values of charges for
which we want to compute $S_{BPS}$. Then we have the relation:
\ben \label{eindsub}
S_{BPS}(P,Q) &=& S_{\rm classical} +\Delta\ln Z_n
\nonumber \\
&-& 3\, \ln\, L +{3\over 2} \ln{\beta\over M_{BPS}} -{1\over 2}  \ln \det {\p^2 \ln Z_n\over \p\nu_i\p\nu_j}\, ,
\een
where  $\Delta\ln Z_n$ is the one loop
correction to $\ln Z_n$, $L$ is the size of the box (in string units) 
in which we place the whole system, $S_{\rm classical}$ denotes the
classical contribution to the entropy given by Wald's formula\cite{9307038} and 
$\nu_i=\ointop d\tau A^{(i)}_\tau$
where $A^{(i)}_\mu$'s are the U(1) gauge fields and $\ointop$ denote integration over the 
Euclidean time circle parametrized by $\tau$. 
The terms in the second line of \refb{eindsub} arise from having to change ensemble from
variable electric charges and momentum to fixed charges and momentum.
As in \cite{2605.30417}, 
we shall take the asymptotic metric to be
$\beta^2\eta_{\mu\nu}$ so that $\tau$ has fixed period 1. In this coordinate system, if we scale all the
charges by $\alpha$, the metric scales as $\alpha^2$.

If $\phi$ denotes the dilaton field and 
$g_s$ denotes the string coupling, given by the value of $e^{\phi}$ in the black hole background,
then we have
\be\label{ensdep}
\det {\p^2 \ln Z_n\over \p\nu_i\p\nu_j} = g_s^{-2 n_{NS}}\, ,
\ee
where $n_{NS}$ is the number of $U(1)$ gauge fields arising in the NSNS sector of the theory.
As discussed in \cite{2605.30417},
the asymmetry between the NSNS and the RR sector arises due to the fact that  the kinetic
terms of the NSNS sector gauge fields carry a factor of $e^{-2\phi}\sim g_s^{-2}$ but the kinetic terms
of the RR sector gauge fields do not carry any such factor\cite{Polchinski:1998rr}. Formally, this
follows from the scaling law described in the second paragraph of the introduction.

The Schwarzschild radius of the black hole is of order $G_N M_{BPS}$ and the Newton's constant
$G_N$ is of order $g_s^{2}$ in string units. On the other hand $\beta$ has been taken to be
of the order of the Schwarzschild radius. Hence we have\footnote{If we had chosen to represent all
lengths and mass in Planck units then $\beta/M_{BPS}$ will be of order unity but $L$ in \refb{eindsub} will
represent the size of the box measured in Planck scale. The combination appearing in \refb{eindsub} will
remain unchanged.}
\be
{\beta\over M_{BPS}}  \sim g_s^2\, .
\ee

The contribution to $\Delta \ln Z_n$ involving logarithms of large arguments comes from various sources.
First of all we have contribution from non-zero eigenvalues of the kinetic operator. In string theory
where string scale provides the natural cut-off, this contribution takes the form
\be\label{enz}
\alpha_{NZ}\ln (\lambda_S)\, ,
\ee
where $\alpha_{NZ}$ takes different values in different theories:
\be \label{ealphans}
\alpha_{NZ}=
\cases{
n-15 \quad \hbox{for $\mathcal N=8$ theory} \, ,\cr
n-11 \quad \hbox{for $\mathcal N=6$ theory} \, ,\cr
n-7 \quad \hbox{for $\mathcal N=4$ theories} \, , \cr
{1\over 6} \, (11 - n_V + n_H)
 - (5-n)\quad \hbox{for $\mathcal N=2$ theories}\, .
}
\ee
Note that since $2n$ is the number of broken supersymmetries, we have $n=14$, 
$n=10$, $n=6$ and $n=2$
respectively for the $\mathcal N=8$, $\mathcal N=6$, 
$\mathcal N=4$ and $\mathcal N=2$ theories but we have not substituted these values.
For the $\mathcal N=2$ theories, $n_V$ and $n_H$ are respectively the numbers of vector and hypermultiplets. The result for the $\NN=6$ theory was not given in \cite{2605.30417} 
but we have derived it
in appendix \ref{sa}.

Contribution to $\Delta \ln Z_n$ from the three translational zero modes take the form:
\be\label{etrans}
\ln (g_s^{-3} L^3 \lambda_S^3)\, .
\ee
Contribution to $\Delta \ln Z_n$ from the two rotational zero modes take the form:
\be\label{erot}
\left( g_s^{-1} \lambda_S^2 \right)^2\, .
\ee
Contribution to $\Delta \ln Z_n$ from the $2n$ fermionic zero modes associated with broken
supersymmetry take the form:
\be\label{eferm}
\ln \left(\lambda_S^{-n}\right)\, .
\ee
Finally the contribution to $\Delta \ln Z_n$ from the two Kalb-Ramond 2-form zero modes take
the form
\be\label{ekalb}
\ln (g_s^{-2})\, .
\ee
There are also possible zero modes of RR 2-form fields but it was shown in 
\cite{2605.30417} they do not
give any $g_s$ or $\lambda_S$ dependent terms.\footnote{Note that 
the only $p$-form fields with normalizable zero modes in 
the Euclidean black hole background are the 2-form fields. So we do not need to worry about the
zero modes of other RR $p$-form fields.}
In fact, we shall argue in section \ref{s4} that the
zero modes of RR fields never contribute any background dependent term  
in the string (field) theory path integral.

Adding all the contributions, we get the following expression for the logarithmic correction to the index
\be\label{etotal}
\Delta S_{BPS} = (\alpha_{NZ}+7-n) \ln\lambda_S + (n_{NS}-4) \ln g_s\, .
\ee

\sectiono{Logarithmic corrections in pure D-brane system} \label{s3}

In our analysis we shall consider three classes of theories. The first one involves type IIA / IIB string theory 
compactified on a six dimensional torus. Since the two theories are related by T-duality, we can focus on the
type IIA description. In this theory we can consider a system carrying only D-brane charges, e.g. D6 wrapped on
$T^6$ and D2-branes wrapped on the three orthogonal planes (45, 67 and 89), which turns into a black hole
when gravitational backreaction is taken into account. We shall denote by $\lambda$ the common scale of the 
D-brane charges, measured in integer units, 
carried by the black hole. The entropy of the black hole is known to scale as $\lambda^2$. Hence
the horizon size, measured in the Planck scale, scales as $\lambda$. Another feature of these black holes is
that the dilaton does not flow in this geometry and can be set to any constant value. One way to see this is to
note that there is a consistent truncation of this theory to an $\mathcal N=2$ supersymmetric theory in which we project
out all fields that are not invariant under simultaneous $2\pi/3$ rotation in the 45, 67 and 89 planes. In this theory
the dilaton is a field in the hypermultiplet and hence does not flow in the black hole background. 
The second theory we consider is the $\NN=6$ theory, obtained by starting with type IIA string
theory on $T^4\times T^2$ and taking a quotient of this  by the $\ZZZ_2$ orbifold group that acts as a shift along $T^2$
and as a T-duality along all the circles of $T^4$. 
In this theory we can construct black holes carrying only
D-brane charges by combination of D6-branes along $T^4\times T^2$, 
D2-branes along $T^2$ and D2-branes along
the two $T^2$'s inside
$T^4$, with their numbers adjusted so that the whole configuration is invariant under the orbifold group.
The entropy scales as the square of the scale $\lambda$ of the D-brane charges and the string coupling
$g_s$ can take arbitrary value in this background.
The third
class
of theories we consider are $\mathcal N=4$ supersymmetric theories, including type II theory of $K3\times T^2$ and its
various orbifolds that involve a  translation along a circle of $T^2$ and an action on $K3$ that preserves
all the space-time supersymmetries of the original theory. In this theory also we can construct systems carrying only
D-brane charges e.g. D6-branes wrapped on $K3\times T^2$, and D2-branes wrapped on $T^2$ and 2-cycles of $K3$
that are preserved by the orbifold group action. For sufficiently generic charges, 
these configurations turn into black holes when gravitational backreaction is taken into
account. The entropy still scales as $\lambda^2$, -- the square of the scale of D-brane charges and the string coupling
can still take arbitrary constant value $g_s$ in the background. The final systems we shall consider are type IIA and
IIB string theories on Calabi-Yau threefolds. This also has system of D-branes that turn into black holes when
gravitational backreaction is taken into account, with the entropy scaling as the 
square of the D-brane charge $\lambda$
and the string coupling taking arbitrary value $g_s$.

In all of these systems the horizon size of the black hole measured in Planck units is of order $\lambda$ and the string
coupling is $g_s$. 
Therefore the horizon size measured in the string scale is given by
\be\label{elambdap}
\lambda_S = g_s\lambda\, .
\ee
We shall take 
$\lambda_S$ to be large  and $g_s$ to be small. In this case both
higher derivative and string loop corrections are under control and we can trust perturbation theory. Our goal will
be to check if the one loop correction to the logarithm of the index, given in \refb{etotal}, is independent of
$g_s$. This is necessary for consistency, since $g_s$ is a free parameter and the index should not depend
on it.

Substituting \refb{elambdap} into \refb{etotal} we get\footnote{The $\ln g_s$ terms coming from
the path integral are closely
related to the ones analyzed in \cite{1002.3805,1810.11343} except that 
here we have both an infrared cut-off provided
by the black hole size and an ultraviolet
cut-off provided by the string scale.}
\be\label{etotalb}
\Delta S_{BPS} = (\alpha_{NZ}+7-n) \ln\lambda + (\alpha_{NZ}+3-n + n_{NS}) \ln g_s\, .
\ee
Now for $\mathcal N=8$ theory corresponding to type II on $T^6$, we have altogether 12 NS sector gauge fields
from the components of the metric and the Kalb-Ramond field along the six circles. 
For the $\NN=6$ theory this number is eight, since all four NS sector gauge fields associated with the
$T^2$ direction and half of the eight NS sector gauge fields associated with the $T^4$ direction
survive the orbifold projection.
For type II string theory
on K3$\times T^2$, this number is four, originating from the components of the metric and the
Kalb-Ramond field along the two circles. Finally, for type IIA and IIB on Calabi-Yau three folds, the number
of NS sector gauge fields is zero since there are no circle directions. Thus we have
\be\label{enns}
n_{NS} = \cases{12 \quad \hbox{for $\mathcal N=8$ theory} \, ,\cr
8 \quad \hbox{for $\mathcal N=6$ theory} \, ,\cr
4 \quad \hbox{for $\mathcal N=4$ theories} \, , \cr
0\quad \hbox{for $\mathcal N=2$ theories}\, .
}
\ee
Substituting this and the value of $\alpha_{NS}$ from \refb{ealphans} into \refb{etotalb}, we get
\be\label{esbps}
\Delta S_{BPS} = \cases{-8\ln\lambda \quad \hbox{for $\mathcal N=8$ theory} \, ,\cr
-4\ln\lambda \quad \hbox{for $\mathcal N=6$ theory} \, ,\cr
0 \quad \hbox{for $\mathcal N=4$ theories} \, , \cr
{1\over 6} \, (23 - n_V + n_H) \ln\lambda
 +{1\over 6} \, (-1 - n_V + n_H)\ln g_s \quad \hbox{for $\mathcal N=2$ theories}\, .
}
\ee
From this we see that the dependence on $g_s$ indeed disappears for the $\mathcal N=8$,
$\NN=6$ and $\mathcal N=4$ theories
as desired. For the $\mathcal N=2$ theories the coefficient of $\ln g_s$ is given by
\be\label{eeuler}
{1\over 6} \, (-1 - n_V + n_H) =\pm {\chi\over 12}\, ,
\ee
where $\chi$ is the Euler number of the Calabi-Yau manifold and the $+/-$ signs are for type IIB/IIA 
theories. Thus we see that the $g_s$ dependence disappears when the underlying Calabi-Yau space has
Euler number zero but not otherwise. We shall discuss this further in section \ref{s5} but will not arrive
at any concrete explanation.

We note in passing that if we had considered black holes carrying pure NSNS
charges then in \refb{etotal} 
$\lambda_S$ would be given by the scale $\lambda_P$ of the magnetic charges and $g_s$ will be
fixed at $\sqrt{\lambda_P/\lambda_Q}$ where $\lambda_Q$ is the scale of the electric 
charges\cite{2605.30417}.
Hence from \refb{etotal} the logarithmic correction to the entropy of such black holes will be given by,
\be\label{esbpsnew}
\Delta S_{BPS}' = \cases{-8\ln\lambda_P + 4\ln {\lambda_P\over \lambda_Q}
\quad \hbox{for $\mathcal N=8$ theory} \, ,\cr
-4 \ln\lambda_P + 2\ln {\lambda_P\over \lambda_Q} \quad \hbox{for $\mathcal N=6$ theory} \, ,\cr
0 \quad \hbox{for $\mathcal N=4$ theories} \, .
}
\ee
There are no results for $\NN=2$ theory since $n_{NS}=0$ for these theories. Also note that
for the $\NN=4$ theory the result differs from that in \cite{2605.30417}, but this is not surprising since we are
in a different duality frame and large $\lambda_Q/\lambda_P$ here does not correspond to 
large $\lambda_Q/\lambda_P$ in \cite{2605.30417}. In fact the results for $\NN=8$ and $\NN=4$ theories
can be shown to be consistent with the known microscopic results for the index in these theories.

\sectiono{Dependence on the ultraviolet cut-off}  \label{s4}

The analysis described above assumes that the intrinsic ultraviolet cut-off of the theory is the string
scale. This is the reason why in \refb{enz} the argument of the logarithm is the horizon size measured
in the string scale. This input has also been used in evaluating the various zero mode contributions
in \refb{etrans}, \refb{erot}, \refb{eferm} and \refb{ekalb}. Our goal in this section will be to examine how the
result changes when we use a different ultraviolet cut-off. In particular we shall consider the case where the
ultraviolet length cut-off is of order $g_s^\gamma$ times the string length scale.

As discussed in appendix \ref{sb},
when the ultraviolet cut-off is the string scale then the path integral over a string field component 
$C$ (ignoring higher order terms in the action) takes the
form
\be\label{emeas}
\int \prod_x dC(x) \, \exp\left[{1\over 2} \int d^4 x \, C(x) \, G^{\mu\nu} \p_\mu \p_\nu C(x)\right]\, ,
\ee
where $G^{\mu\nu}$ is the inverse string metric.
In this formula we have not displayed any tensor index that $C$ might carry. All such tensor indices
must be contracted with flat metric. Also any $\sqrt{\det G}$ and dilaton / other scalar field
dependent factors that might have been present originally
have been absorbed into $C$. More detailed discussion on this and its relation to ultraviolet cut-off can
be found in appendix \ref{sb}. A string field theory derivation of \refb{emeas} can be found in
\cite{2605.30417}.

Since $\lambda_S$ is the overall size of the black hole in string metric, we see that the eigenvalues
of $G^{\mu\nu} \p_\mu \p_\nu$ are of order $1/\lambda_S^2$. Hence integration over each mode produces
a factor of $\lambda_S$. This is the reason why the argument of the logarithm
\refb{enz} is $\lambda_S$. Since the choice
of ultraviolet cut-off is correlated with the normalization of the path integral measure, the integral over the
zero modes also depends on the choice of ultraviolet cut-off. This is implicit in the results given in
\refb{etrans}, \refb{erot}, \refb{eferm} and \refb{ekalb}, but we shall see this more explicitly below.

Now suppose that we want to change the ultraviolet cut-off length to $g_s^\gamma$ times the string length.
For example, if the cut-off had been set by the Planck length then we would have 
$\gamma=1$.
It follows from the discussion in appendix \ref{sb} that  in \refb{emeas} we need to replace the string metric
$G_{\mu\nu}$ by a new metric $\GG_{\mu\nu}$ such that when measured in the metric $\GG_{\mu\nu}$ the
ultraviolet cut-off length is of order unity. This gives
\be
\GG_{\mu\nu} = g_s^{-2\gamma} G_{\mu\nu}\, .
\ee
Hence \refb{emeas} should be replaced by
\be\label{emeas1}
\int \prod_x dC_{new}(x) \, \exp\left[{1\over 2} \int d^4 x\, C_{new}(x) \, g_s^{2\gamma} G^{\mu\nu} 
\p_\mu \p_\nu C_{new}(x)\right]\, ,
\ee
The eigenvalues of the kinetic operator are now given by $g_s^{2\gamma} \lambda_S^{-2}$ instead of
$\lambda_S^{-2}$. Hence $\lambda_S$ is replaced by $g_s^{-\gamma}\lambda_S$. For half integer spin fields
the original kinetic operator was $E_a^\mu \p_\mu$ where $E^a_\mu$ is the inverse vierbein constructed from the
inverse string metric $G^{\mu\nu}$ and the eigenvalues of the kinetic operator were
of order $\lambda_S^{-1}$. Now the
kinetic operator will be $g_s^{\gamma} E_a^\mu \p_\mu$ and the eigenvalues will be of order 
$g_s^\gamma\lambda_S^{-1}$.
Hence again the net effect is to replace $\lambda_S$ by $g_s^{-\gamma}\lambda_S$. Hence the net non-zero mode
contribution given in \refb{enz} changes to
\be\label{enzgamma}
\alpha_{NZ}\, (\ln\lambda_S - \gamma \ln g_s)\, .
\ee

We now turn to the contribution of the zero modes. For this note that the action appearing in the exponent
of \refb{emeas1} can be mapped to that in the exponent of \refb{emeas} if we identify
\be\label{eccnew}
C_{new} = g_s^{-\gamma} \, C\, .
\ee
However since the integration variable is still $C_{new}$, we shall get an extra factor of $g_s^{-\gamma}$
for each bosonic mode, including the zero modes. 
The non-zero mode integrals have already been analyzed in \refb{enzgamma}, 
so the main new result is that each bosonic
zero mode integral will produce an extra factor of $g_s^{-\gamma}$.
For fermionic zero modes there are two differences. 
First of all,
$g_s^{-\gamma}$ in \refb{eccnew} is replaced by $g_s^{-\gamma/2}$ since the kinetic term is multiplied
by $g_s^\gamma$ instead of $g_s^{2\gamma}$. Second, due to the grassmann odd nature of the fermions,
each fermion zero mode integral will now produce a factor of $g_s^{\gamma/2}$ instead of
$g_s^{-\gamma/2}$. Taking into account the fact that we have altogether seven bosonic zero modes
including translation, rotation and 2-form zero modes and $2n$ fermion zero modes, we get a net
contribution of
\be \label{ezerogam}
g_s^{-7\gamma + n\gamma}\, .
\ee
Note that we have not included possible $\gamma$ dependent 
contribution from the zero modes of the RR sector 2-form
fields. We postpone this discussion till the end of this analysis.
Combining \refb{ezerogam} with the non-zero mode contribution \refb{enzgamma} we see that the net extra
contribution to $\ln Z_n$ due to the change of cut-off scale is given by
\be \label{egammadep}
(-\alpha_{NZ}-7+n)\, \gamma\, \ln g_s\, .
\ee
Using \refb{ealphans} we can express this as\footnote{Note that the coefficient of $\gamma\ln g_s$ is
given by the full conformal anomaly of the theory including the zero mode contribution. Hence it can be
computed using the full heat kernel\cite{0306138} after replacing the NSNS sector axion by the 2-form field
following \cite{Duff:1980qv}.}
\be \label{ecorrection}
\cases{
8\,  \gamma\, \ln g_s \quad \hbox{for $\mathcal N=8$ theory} \, ,\cr
4\,  \gamma\, \ln g_s \quad \hbox{for $\mathcal N=6$ theory} \, ,\cr
0 \quad \hbox{for $\mathcal N=4$ theories} \, , \cr
{1\over 6} \, ( n_V - n_H-23 ) \,  \gamma\, \ln g_s
 \quad \hbox{for $\mathcal N=2$ theories}\, .
}
\ee
This shows that while $\gamma$ dependence cancels in the case of $\NN=4$ theory, it takes
non-zero value for $\NN=8$, $\NN=6$ and $\NN=2$ theories. Hence a different choice of cut-off would spoil
the desired property of the index, namely $g_s$ independence,  for the $\NN=8$ and $\NN=6$ 
theories and $\NN=2$
theories for $\chi= 0$. We also note that by adjusting $\gamma$ we cannot restore the desired
property of the $\mathcal N=2$ theory for $\chi\ne 0$ since the coefficient of $\gamma$ is not proportional
to $\chi$ (unless we make $\gamma$ itself dependent on $\chi$ which will be highly artificial).

Let us now turn to the possible contribution from RR sector 2-form zero modes. Such zero modes
arise from pure gauge configurations with non-normalizable gauge parameters. However in
the string field theory action and its low energy limit,
the interacting part of the theory only involves RR field strengths and
not RR gauge fields\cite{1508.05387,1511.08220,2405.19421,2506.00120}. The latter arise from the $(-3/2,-3/2)$ 
picture vertex operators which only enter the free part of the action. As a result the result of integration
over the RR zero modes cannot be sensitive to the background dilaton or other fields and must
give $g_s$ independent results. For this reason it safe to ignore the contribution from the integration
over the RR zero modes.  

It is instructive to consider the $\gamma=1$ case that corresponds to choosing the Planck scale as the
ultraviolet cut-off. In this case adding \refb{egammadep} to the $g_s$ dependent contribution in
\refb{etotalb} we get the net $g_s$ dependent contribution:
\be\label{etotalc}
 (n_{NS}-4) \ln g_s\, .
\ee
Both the terms in this expression can be traced to the charge quantization laws. For example the
$n_{NS}$ term can be traced to the fact that the NSNS sector gauge fields that couple to the integer
quantized charges have a non-canonical kinetic term with $g_s^{-2}$ multiplying it. The $-4$ can be traced to
a similar observation on the Kalb-Ramond 2-form that couples to the integer quantized charge namely the
fundamental string. These cannot be removed by changing the normalization of the gauge fields and $B_{\mu\nu}$
since $g_s^{-2}$ is the vacuum expectation value of the dilaton field $e^{-2\phi}$ and absorbing $\phi$ dependent
factors into the definition of the gauge fields will lead to non-standard gauge transformation laws and non-standard
conservation laws for the charges they couple to. 
This fits in with the comment in the second paragraph of the introduction that the
information on the string scale enters through the charge quantization laws.

Finally, let us note that in this analysis we have assumed that as we change the ultraviolet cut-off we only
change the cut-off and nothing else. By relaxing this in a coordinated manner we could get $g_s$ independence
for non-zero $\gamma$. 
For example, we could work in different duality frames in appropriate limits of the
asymptotic values of moduli fields and in each of these frames the cut-off scale is different,
although at the end the dependence of the asymptotic values of the moduli fields must
disappear.
As an example, we can study the strong coupling limit where we can use S-duality to argue that the effective
cut-off will be given by the length of the dual string. This is $g_s^2$ in the original string units and hence
corresponds to the case $\gamma=2$. So we can use \refb{egammadep} to study this case. However,
there are several other effects to be taken into account. The first effect is that the duality does not
act simply on the Kalb-Ramond two form, it acts on the dual axion locally. So we remove the
Kalb-Ramond 2-form zero modes from the analysis of $\gamma$ dependence and use \refb{egammadep}
for $\gamma=2$ as
\be \label{egammadepnew}
(-\alpha_{NZ}-5+n)\, 2 \, \ln g_s\, ,
\ee
where the $-7$ in \refb{egammadep} has changed to $-5$ since instead of seven bosonic zero modes we
are accounting for 5 bosonic zero modes.  But we now need to take into account the effect of the zero modes
of the new Kalb-Ramond 2-form.
The simplest way to do this is to use duality symmetry to replace the $g_s^{-2}$ in \refb{ekalb} by
$g_s^2$. This gives a net extra term in $\ln Z_n$ of the form: 
\be \label{eadd1}
\ln\, g_s^4 = 4\, \ln g_s\, .
\ee
Finally, in the strong coupling limit the roles of electric and magnetic charges get exchanged.
As a result the factor of $g_s^{-2n_{NS}}$ in \refb{ensdep} is replaced by $g_s^{2n_{NS}}$.
Hence in \refb{eindsub} we get a net extra term
\be \label{eadd2}
\ln\, g_s^{-2 n_{NS}} = -2\, n_{NS} \, \ln g_s\, .
\ee
Adding \refb{egammadepnew}-\refb{eadd2} we get
\be
2\, (-\alpha_{NZ}-3+n - n_{NS})\, \ln g_s\, .
\ee
Adding this to \refb{etotalb} has the net effect of changing the sign of the $\ln g_s$ term.
Since the coefficient of $\ln g_s$ was zero before, it continues to remain zero.\footnote{Some
of the cancellations would look less mysterious if we had used Planck units instead of string units
to express various quantities, but the final result will remain unchanged.}

\section{Discussion} \label{s5}

In this paper we have shown that in $\NN=8$, $\NN=6$ and $\NN=4$ supergravities in four
space-time dimensions, the supersymmetric index of black holes carrying purely RR charges is independent
of the string coupling as required by supersymmetry if the ultraviolet cut-off of the theory is taken to be
of the order of the string scale. The same result holds for $\NN=2$ supersymmetric string theories
associated with type IIA or IIB string theories on Calabi-Yau manifolds of zero Euler number. On the
other hand for $\NN=8$, $\NN=6$ and $\NN=2$ theories this results fails if the ultraviolet cut-off is taken to be different
from the string scale. This provides strong evidence that in order to make sense of supergravity theories with
extended supersymmetry as quantum theories, we must embed them in superstring theory.

Clearly the most important open question is to understand the case of compactification of type IIA / IIB theory
on Calabi-Yau manifolds of non-zero Euler number $\chi$. In this case the index computed from the
path integral over massless fields, instead of being independent of the
string coupling $g_s$, is proportional to $g_s^{\pm\chi/12}$ where the $+/-$ sign holds for type IIB/IIA
compactification. We do not have an explanation for this discrepancy, but we shall end with a few observations. We divide these into two kinds: the first set will try to identify possible errors in the
path integral calculation and the second set will try to identify possible errors in the microscopic 
calculation. Both sets are arranged in the order of increasing likelyhood.

We first consider possible sources of error in the macroscopic calculation.
\begin{enumerate}
\item We have not included the contribution from massive string states in our analysis. However their effect
can be taken into account as additional local terms in the effective action with a regular power series
expansion in $g_s$ and would not lead to terms proportional to $\ln g_s$.
\item When the Euler number of the Calabi-Yau threefold is non-zero, there are many new terms in the effective
action including a renormalization of the Einstein-Hilbert term\cite{9707013}. 
However these corrections can be reinterpreted as correction to the hypermultiplet moduli
space metric\cite{0602164,1009.3026,1010.5792} and
does not seem to introduce any $g_s$ dependence of the index. 
There are various other contexts in which the Euler number of the Calabi-Yau threefold
has appeared in the analysis  of type II compactification
(see {\it e.g.} \cite{2305.19916,2510.23722}), 
but none of these seem to be directly related to the issue of
$g_s$ dependence of $B_n$ that we are finding.
\item If we consider the torus partition function of the world-sheet CFT with periodic boundary conditions
on the fermions along both circles of the torus, then the partition function associated with the Calabi-Yau
sigma model will be given by the Euler number $\chi$ of the Calabi-Yau manifold. 
If we ignore the presence of RR flux background, then, since the black hole space-time has non-zero
Euler number, the partition function of the sigma model with black hole target space will also be 
non-zero.
On the other hand in this
sector we have bosonic zero modes from the $\beta\gamma$ ghost sector and the integration over these
zero modes would diverge making the total partition 
function diverge.\footnote{I wish to thank 
Mukund Rangamani for discussion on
this.}
In the Hamiltonian formulation this
divergence can be traced to the infinite degeneracy of the Ramond sector ground state in the 
$-1/2$ picture created by the zero mode of $\gamma$..
This type of problem has already been observed in
\cite{2408.14567,2505.14380,mslg} in a different context.
This problem is absent for $\chi=0$ since the matter sector partition function
vanishes and we can regulate the ghost and the matter partition function by introducing a twisted boundary
condition and taking the twist parameter to zero at the end of the computation. It is also absent in
$\NN=4$, $\NN=6$ and $\NN=8$ theories due to the presence of fermion zero modes in the matter sector associated
with the compact circle directions.

One can resolve this problem in string (field) theory by computing the one point function of various fields, for
which the appropriate insertion of picture changing operators prevent the infinitely degenerate ground states
in the $-1/2$ picture to propagate, and then integrating this to find the partition function. Nevertheless
the divergence in the original partition function suggests that the scaling argument that we have used
to determine the $g_s$ dependence of the partition function may not quite be valid when $\chi$ is
non-vanishing. In particular if this effect generates an extra 
term in the effective action of the form $\chi \, \phi$ times
the four dimensional 
Euler density with appropriate coefficient, then this can generate a term of the form 
$g_s^{\mp\chi/12}$ in the partition function and cancel the unwanted term.
This clearly deserves a more detailed analysis, taking into account the effect of the
background Ramond Ramond fluxes.
\end{enumerate}
We now try to identify possible errors in the microscopic computation.
\begin{enumerate}
\item 
If there are no errors in the computation of $B_n$ from string field theory path integral, then we need to explore the
possibility that at finite temperature, $B_n$ receives contribution other than that from single centered black holes
which could explain its dependence on $g_s$. Indeed,
it was noted in several papers\cite{1406.2360,1501.01643,1808.05606} that at finite temperature, the index
$B_{n}$ of the type we are considering can receive contribution not only from single centered black hole but also
scattering states involving multi-centered black holes. 
The latter contribution is temperature dependent. However this contribution remains independent of the
hyper-multiplet moduli including the dilaton, and cannot explain the observed $g_s$ dependence.
\item The analysis of \cite{1406.2360,1501.01643,1808.05606} focussed on the contribution of multi-particle
states where each of the particles is a charged black hole. However in principle we could also have contribution
from multiparticle states where one of the particles is the original black hole 
and the others are
massless states
belonging to the gravity multiplet, vector multiplet and the hypermultiplet. The gravity multiplet and each vector multiplet
contributes 2 to the $B_2$ index while the hypermultiplet contributes $-2$ to the $B_2$ index. 
Therefore the total contribution to $B_2$ from the massless states will be given by
$2(n_V+1-n_H)=\mp\chi$ as in \refb{eeuler}. If the contribution comes from states involving the original
black hole and arbitrary number of massless states of this type, then the contribution will naturally exponentiate,
producing a factor of the form $e^{c\chi}$ for some constant $c$. 
While this remains a plausible explanation, it remains to be seen if this can really be used to give a 
fully consistent explanation of the $g_s^{\pm\chi/12}$ factor.

If this is the explanation of the $g_s$ dependence of $B_n$, then it
would also explain why such $g_s$ dependence did not appear for $\NN=8$, $\NN=6$ 
and $\NN=4$ supersymmetric theories. In these theories
the generic multi-centered contribution to the index vanishes due to additional fermion zero modes carried
by individual centers\cite{0702141,0903.2481,0803.1014}, leaving behind at most exponentially suppressed
multi-centered contributions.
\end{enumerate}
It will also be interesting to explore whether the issues we encountered have any relation with the
apparent difficulty in matching the microscopic and macroscopic results for the logarithmic terms in
the five dimensional string theory\cite{2605.19552}.
We hope to return to a fully satisfactory resolution of these issues in the future.

\bigskip

\noindent{\bf Acknowledgement:} 
I wish to thank Muktajyoti Saha and P. Shanmugapriya for discussions
and collaboration in \cite{2605.30417} 
whose results
were used in this paper. I also thank 
Sergey Alexandrov, Boris Pioline and Mukund Rangamani for useful communications on possible
origin of the $g_s$ dependence of the index in type II string theories compactified on Calabi-Yau
manifolds of non-zero Euler number.
This work was supported by the ICTS-Infosys Madhava 
Chair Professorship and
Department of Atomic Energy, Government of India, under project no.~RTI4019.

\appendix

\sectiono{Computations in the $\NN=6$ supersymmetric theory} \label{sa}

In this appendix we shall derive the result \refb{ealphans} 
for $\NN=6$ supersymmetric theory in four dimensions.
For this we shall use the result of \cite{1109.0444} that computes the total logarithmic correction to
the black hole index in $\NN=6$ supersymmetric theory when all charges are large of order $a$,
and string coupling is of order unity. This is
given by
\be \label{efulln6}
-4 \, \ln a\, .
\ee
This result was based on the results from the near horizon analysis, but as shown in \cite{2306.07322} 
the same result
can be obtained from the computation of the $B_{10}$ index in the full space-time geometry.
We now need to remove the zero mode contribution from this. \refb{efulln6} includes contribution
from the three translation zero modes giving a net contribution $3\ln a$, 
two rotation zero modes giving a net contribution of $4\ln a$ and
$2n$ gravitino zero modes (here $2n=20$) giving a net contribution of 
$-n\ln a$.\footnote{These can be
read out from \refb{etrans}-\refb{eferm} with $\lambda_S$ replaced by $a$ and ignoring the other factors.}
Subtracting these from \refb{efulln6} we get
the net non-zero mode contribution to the logarithmic correction to be
\be
(-4-3-4+n)\ln a=(n-11)\ln a\, .
\ee
Finally we use the fact that the non-zero mode contribution to the logarithmic correction should
involve the size of the black hole measured in units of ultraviolat cut-off that we are taking to
be the string scale. Hence $a$ is replaced by $\lambda_S$ and we get
\be
(n-11)\ln \lambda_S\, .
\ee
This reproduces \refb{ealphans}.

\sectiono{Ultraviolet cut-off and path integral measure} \label{sb}

It is generally stated that the one loop contribution to the effective action due to a bosonic field is
given by the square root of the inverse of the determinant of the kinetic operator of the field.
This statement needs some qualification. Let $C_{a_1\cdots a_n}$ be a multi-component field, normalized
so that its path integral measure is of the form
\be\label{epath}
\prod_x dC_{a_1\cdots a_n}(x)\, ,
\ee
without any other factor. Here the $a_i$'s can be tensor indices or some internal indices.
Let us further suppose that the Euclidean action takes the form
\be\label{ekin}
{1\over 2} \int d^4 x \, C_{a_1\cdots a_n} \GG^{\mu\nu} \p_\mu\p_\nu C_{a_1\cdots a_n}\, .
\ee
where $\GG^{\mu\nu}$ is some inverse metric. Note in particular that there is no metric factor involved
in contracting the indices of $C_{a_1\cdots a_n}$ even if the $a_i$'s are tensor indices
and there is also no factor of $\sqrt{\det\GG}$ or any other
scalar, e.g. the dilaton; all such factors
have been absorbed into the definition of $C_{a_1\cdots a_n}$ and 
$\GG^{\mu\nu}$.\footnote{This makes the action looks non-invariant under 
general coordinate transformation, but in
string field theory the invariance is restored by higher order terms in the action. Equivalently one
could use an appropriate ultra-local measure so that the result of integration gives
$\left[\det  \GG^{\mu\nu}\p_\mu\p_\nu\right]^{-1/2}$.}
In that
case the path integral over each $C_{a_1\cdots a_n}$ will indeed produce:
\be\label{eB.3}
\left[\det  \GG^{\mu\nu}\p_\mu\p_\nu\right]^{-1/2} = \exp\left[-{1\over 2} \ln \det  \GG^{\mu\nu}\p_\mu\p_\nu
\right]\, .
\ee

We now use the identity,
\be\label{eproper}
\int_\eps^\infty {ds\over s} e^{-s\Lambda} = -\ln(\Lambda\eps) + \hbox{constant}, 
\qquad \hbox{for $\Lambda\eps
<<1$}\, .
\ee
Our interest will be in applying this formula to the case where $\Lambda$ is an
eigenvalue of the Laplace operator. The condition $\Lambda\eps<<1$ tells us that the formula is
valid only for $\Lambda<<1/\eps$, i.e.\  $\eps$ acts as an ultraviolet cut-off. Since $\Lambda$
has the dimension of the inverse length$^2$, $\eps$ has the dimension of length$^2$ which
can be thought of as the square of the ultraviolet length cut-off. Now, if $\Lambda$ is directly the
eigenvalue of the operator $\GG^{\mu\nu}\p_\mu\p_\nu$, then we need to get $-\ln\Lambda$ on the right hand
side of \refb{eproper} in order to produce the term in the exponent on the right hand side of \refb{eB.3}.
Hence we must have $\eps\sim 1$, i.e.\ the ultraviolet
cut-off measured in the metric $\GG_{\mu\nu}$ should be of order unity. In this way the choice of the
integration measure over the fields $C_{a_1\cdots a_n}$ fixes the metric $\GG^{\mu\nu}$,
and this in turn fixes the choice of the ultraviolet cut-off.

We can consider the example of string theory. The one loop amplitude involves integration over 
the modular parameter $\tau=\tau_1+i\tau_2$ and we can identify $\tau_2$ as $s$. The lower cut-off
on $\tau_2$ is of order unity. $\Lambda$ is the eigenvalue of the $L_0$ operators. For massless states
$L_0$ is given by $G^{\mu\nu}\p_\mu\p_\nu$ up to a numerical factor where $G_{\mu\nu}$ is 
the string metric. Comparing this with the previous analysis we see that the ultraviolet cut-off 
measured in the string metric is of order unity, which is indeed the correct conclusion. 

Independently of this analysis one can show that the correct path integral variables 
$C_{\mu_1\cdots \mu_n}$ of string field theory with measure given in \refb{epath}
have kinetic term of the from given in \refb{ekin} with $\GG^{\mu\nu}$ replaced by the
inverse string metric $G^{\mu\nu}$\cite{2605.30417}.

An identical discussion holds for half-integer spin fields, with $\GG^{\mu\nu}\p_\mu\p_\nu$ replaced by
$\EE_a^\mu\p_\nu$, where $\EE_a^\mu$ is the inverse vierbein associated with the inverse metric
$\GG^{\mu\nu}$.

\end{document}